\documentclass[showpacs,prl,letterpaper,floatfix,nobalancelastpage,twocolumn]{revtex4}
\usepackage{dcolumn}
\usepackage{bm}
\usepackage{amsmath}
\usepackage{bm}
\usepackage{color}
\usepackage{float}
\usepackage{graphicx,subfigure}

\def\braket#1{\mathinner{\langle{#1}\rangle}}

\begin{document}


\title{Phonon-dressed Mollow triplet in the regime of cavity-QED}
\author{C. Roy and S. Hughes}
\email{shughes@physics.queensu.ca}
\affiliation{Department of Physics,
Queen's University,
 Kingston, Ontario K7L 3N6, Canada}



\begin{abstract}
We study the resonance fluorescence spectra of a driven quantum dot placed inside a high $Q$
semiconductor cavity and interacting with an acoustic phonon bath. The dynamics is calculated using a time-convolutionless master equation obtained in the polaron frame.
We demonstrate
pronounced spectral broadening of the Mollow sidebands through cavity-emission which,
for small cavity-coupling rates,
increases quadratically with the Rabi frequency. However,  for larger
  cavity coupling rates,
  this broadening dependence is found to be more
 complex.
%
This field-dependent
Mollow triplet broadening is primarily a consequence of the
triplet peaks sampling different parts of the asymmetric phonon bath,
and agrees directly with  recent experiments with
semiconductor micropillars.
 The influence from the detuned cavity photon bath and multi-photon effects is  shown to
 play a qualitatively important role on the fluorescence spectra.


\end{abstract}

\pacs{42.50.Ct, 78.67.Hc, 78.55.-m}

\maketitle

Recent developments in semiconductor cavity systems
exploit
several fields of research, including  semiconductor optics, quantum information 
processing and cavity-QED (quantum electrodynamics). Target quantum sources such as
indistinguishable photons are important in  technological applications~\cite{ates1},
and  require robust physics designs and an understanding of how to
 manipulate dynamical processes at a very fundamental level.
 Unique quantum processes can be obtained when the cavity and quantum dot (QD) are  near the strong coupling regime~\cite{SC:StrongCoupling1,SC:StrongCoupling2}, which introduces quantum mechanical features such as photon antibunching~\cite{press}. Recent experimental studies involving {\em coherent} excitation have
  focussed on resonance fluorescence of a QD coupled to a cavity mode~\cite{muller,flagg,vamivakas}. Also of interest is the study of off-resonant QD cavity systems that can be used to monitor resonance fluorescence of the emitter~\cite{ates2}. Very large optical dipole moments relative to atomic systems and strong interaction effects are the salient features of these QDs. Yet, the very nature of the semiconductor QDs can
  also introduce
  excitation-dependent dephasing and decoherence effects.

In 1969, Mollow demonstrated
 that the fluorescence spectrum of a two-level atom has additional sidebands in addition to the central
 coherent peak when the atom is driven by a strong coherent single-mode field~\cite{mollow,CarmichaelBook1}. This effect can be attributed to the emergence of ``dressed states''---an infinite series of {\em doublets} or Floquet states, whose positions and spectral widths are  well known for large driving fields.
 From both a practical and fundamental level,
 it is interesting to study such well known atomic optics effects
 when a semiconductor QD is strongly driven by a coherent laser field. For pulsed-excited systems,
 excitation-induced dephasing (EID) emerges due to the interaction of the laser-induced dressed states with the underlying acoustic phonon reservoir~\cite{knorr,mogilevtsev1,ramsay1-2}.
  For continuous (cw) excitation, it is not clear what the role of phonon coupling will be,
  as it is often argued that the dephasing processes are due to the ``non-Markovian''
  processes, especially important for short pulses and fast timescales.
  Thus one may expect that cw excitation does not
  suffer much from phonon-induced damping,
  and indeed recent experimental
  works using QD resonance fluorescence~\cite{muller,flagg,vamivakas} suggest negligible influence from the  phonon bath.
  Nevertheless, if a QD system is driven strongly enough,
  a phonon-induced dephasing process may occur~\cite{nazir1}.
%
%
 In recent experiments~\cite{Ulrich:PLR2010}, cw resonance fluorescence emission from single QD placed inside a high-quality micropillar cavity was studied.
 In contrast to all previous works on cw-excited semiconductor systems, it was found that increasing excitation power results in a  Mollow triplet spectra with a systematic spectral sideband broadening that is proportional to the square of the Rabi frequency---which was interpreted as a signature of EID~\cite{knorr,mogilevtsev1}.
 Since the
 QD was placed in the regime of cavity-QED, which enables significant
 cavity coupling, the origin of this Mollow sideband dephasing
 is not at all clear.


In this work, we develop a nonperturbative theory of
cavity-QED coupling in a longitudinal acoustic (LA) phonon bath. We apply this theory
to connect directly to the EID observations on the Mollow triplet, and we introduce
the phenomenon of:
{\em phonon-dressed  Mollow triplet in the regime of cavity-QED}.
We develop a model
that describes  both electron-acoustic phonon coupling and cavity photon coupling
to all orders.
Specifically, we  describe the process of electron-photon and electron-phonon coupling
by using a polaron transform~\cite{mahan}, which, in the appropriate
limits, formally
recovers the {\em independent boson model}
or the
Jaynes-Cumming model.
The polaron transform allows one to eliminate the QD phonon interaction part from the system Hamiltonian at the expense of modified Rabi frequency, and the dot-cavity coupling that is now {\em dressed} by coherent phonon displacement operators~\cite{mahan,imamoglu}.
We then perform time local projection operations and derive a straightforward time-convolutionless master equation
that treats the residual system-bath interaction to second order~\cite{nazir2}, while
treating cavity photon effects to all orders. {\em To the best of our knowledge, this is the first study
of its kind to explore electron-LA-phonon interactions and cavity photon interactions to all orders.}



Working in a frame rotating with respect to the laser pump frequency,
the model Hamiltonian is
\begin{align}
\label{sec1eq1}
H& =\hbar(\omega_{x}-\omega_{L}){\sigma}^{+}{\sigma}^{-}+\hbar(\omega_{c}
-\omega_{L}){a}^{\dagger}{a} \nonumber \\
&+ \hbar g({\sigma}^{+}{a}+{a}^{\dagger}{\sigma}^{-})+\hbar \Omega({\sigma}^{+}+{\sigma}^{-})   \nonumber \\
&+ {\sigma}^{+}{\sigma}^{-}\sum_{k}\hbar\lambda_{k}({b}_{k}
+{b}_{k}^{\dagger})+\sum_{k}\hbar\omega_{k}{b}_{k}^{\dagger}{b}_{k},
\end{align}
where  ${b}_{k}({b}_{k}^{\dagger})$ are the annihilation and creation operators of the phonon reservoir,
 $a$ is the {\em leaky} cavity mode annihilation operator,  ${\sigma}^+,{\sigma}^-$ (and
 $\sigma^z =\sigma^+\sigma^- - \sigma^-\sigma^+$)  are the Pauli operators of 
the  electron-hole pair (``exciton''),
$\delta_{\alpha L}\equiv \omega_\alpha-\omega_L$ ($\alpha =x,c$)
are the detunings of the exciton and cavity from the coherent pump laser,
 and $g$ is the cavity-exciton  coupling strength.
We next  perform a polaron-transformation that
introduces a renormalized Rabi frequency and dot-cavity coupling strength~\cite{imamoglu}:
$H^{\prime}=\exp(S)H\exp(-S),$
with $S={\sigma}^{+}{\sigma}^{-}\sum_{k}
\frac{\lambda_{k}}{\omega_{k}}({b}_{k}^{\dagger}-{b}_{k}).$
The transformed Hamiltonian has three contributions,
\begin{subequations}
\begin{align}
\label{sec3eq1}
H^{\prime}_{sys} &  =    \hbar(\delta_{xL}-\Delta_{P}){\sigma}^{+}{\sigma}^{-}
+\hbar\delta_{cL} {a}^{\dagger}{a}+\langle B\rangle  X_{g},  \\
H^{\prime}_{bath} & =  \sum_{k}\hbar\omega_{k}{b}_{k}^{\dagger}{b}_{k},\\
H^{\prime}_{int} & =   X_{g}  \zeta_{g}+X_{u}  \zeta_{u},
\end{align}
\end{subequations}
where
$ X_{g} = \hbar g({a}^{\dagger}{\sigma}^{-}+{\sigma}^{+}{a})
+\hbar \Omega({\sigma}^{+}+{\sigma}^{-})$,
$ X_{u} = i\hbar[g({\sigma}^{+}{a}-{a}^{\dagger}{\sigma}^{-})
+\Omega({\sigma}^{-} - {\sigma}^{+})]$,
${B}_{\pm}=\exp[\pm\sum_{k}\frac{\lambda_{k}}{\omega_{k}}({b}_{k}-{b}_{k}^{\dagger})]$,
$ \zeta_{g} =\frac{1}{2}( B_{+}+ B_{-}-2\langle  B\rangle)$, and
$ \zeta_{u} =\frac{1}{2i}( B_{+}- B_{-})$.
Note that $\langle  B\rangle=\langle  B_{+}\rangle=\langle  B_{-}\rangle$ and the polaron shift is given by $\Delta_{P}=\sum_{k}\frac{\eta_{k}^{2}}{g_{k}}$. Without loss of generality,
we assume that the polaron shift is implicitly included in $\omega_{x}$. It is important to note the slightly different definition of the system Hamiltonian in Eq.~(\ref{sec3eq1}). The usual decomposition of the system Hamiltonian to include only the noninteracting QD and cavity parts is known to fail the detailed balance condition in general, which is caused by {\em internal coupling effects}~\cite{carmichael}.
 The  transformed system Hamiltonian above leads to the correct form of the density operator and preserves detailed balance. 

 Next, we introduce  dissipative terms that describe the radiative decay of the QD exciton and the cavity mode. They are included as Liouvillian superoperators acting on the reduced system
 density matrix~\cite{ota}:
\begin{align}
\label{sec2eq34}
L(\rho)& =\frac{\tilde{\Gamma}_{x}}{2}(2{\sigma}^{-}\rho{\sigma}^{+}
-{\sigma}^{+}{\sigma}^{-}\rho-\rho{\sigma}^{+}{\sigma}^{-}) \nonumber \\
& +\kappa(2{a}\rho{a}^{\dagger}-{a}^{\dagger}{a}\rho-\rho{a}^{\dagger}{a})
+ \frac{\Gamma^{\prime}}{4}({\sigma}_{z}\rho{\sigma}_{z}-\rho),
\end{align}
where $2\kappa$ is the cavity decay rate, $\Gamma'$ is the pure dephasing rate,
and  $\tilde{\Gamma}_{x}=\Gamma_{x}\langle  B\rangle^{2}$ is the radiative decay rate.
The latter decay
has an additional  renormalization by a factor of $\langle  B\rangle^{2}$ resulting from the polaron transform. The physical origin of this can be traced to the dephasing of the optical dipole moment resulting from the Franck-Condon displacement of the excited state~\cite{roy}.
In writing the phonon induced renormalization in this form, we have assumed a separation of the timescales associated with the radiative and phonon processes which amounts to replacing the coherent phonon displacement operators ${ B}_{\pm}$ by $\langle B\rangle=\langle{B}_{\pm}\rangle$, where the angular brackets denote trace over the thermal phonon reservoir. This can be justified by noting that phonon dynamics is extremely fast and happens at picosecond timescales, i.e., much faster than the
typical radiative processes which occur.
We now derive
a time-local master equation~\cite{nazir2} using the time-convolutionless approach for the cavity-QED system density operator in the interaction picture, described by $H^{\prime}_{sys}$, and to second order  in exciton-photon-phonon coupling $H_{int}^{\prime}$. The phonon reservoir is considered to be stationary and in thermal equilibrium and factorized initial conditions are assumed.
The time-convolutionless master equation for the cavity-QED system density operator is then given by
\begin{align}
\label{sec3eq3}
\frac{\partial \rho}{\partial t}&=\frac{1}{i\hbar}[H_{sys}^{\prime},\rho(t)]+L(\rho)-\frac{1}{\hbar^{2}}\int^{t}_{0}
d\tau\sum_{m=g,u}  \\ \nonumber
& \times (G_{m}(\tau)[X_{m},e^{-iH_{sys}^{\prime}\tau/\hbar}
X_{m}e^{iH_{sys}^{\prime}\tau/\hbar}\rho(t)]+H.c.) ,
\end{align}
where $G_{g/u}(t)=\langle\zeta_{g/u}(t)\zeta_{g/u}(0)\rangle$. These polaron Green functions are
well known and are defined as~\cite{mahan,imamoglu}
$G_{g}(t)=\langle  B\rangle^{2}(\cosh[\phi(t)]-1)$,
$G_{u}(t)=\langle  B\rangle^{2}\sinh[\phi(t)],$
with
$\phi(t)=\int^{\infty}_{0}d\omega\frac{J(\omega)}{\omega^{2}}[\coth(\beta\hbar\omega/2)\cos(\omega t)-i\sin(\omega t)]$.
The  $\langle  B\rangle$ and the polaron-shift $\Delta_P$ can now be written as
$\langle  B\rangle=\exp(-\frac{1}{2}\int^{\infty}_{0}d\omega\frac{J(\omega)}{\omega^{2}}\coth(\beta\hbar\omega/2))$
and
$\Delta_P=\int^{\infty}_{0}d\omega\frac{J(\omega)}{\omega}$.
This model describes the electron-LA-phonon interaction via a deformation potential coupling which is the primary source of dephasing in self-assembled QDs~\cite{krum}.
%
The spectral function
that considers  electron-phonon interactions via a deformation potential coupling
 can now be written as $J(\omega)=\alpha_{p}\,\omega^{3}\exp(-\omega^{2}/2\omega_{b}^{2})$ where we use $\omega_{b}=1\,$meV and $\alpha_{p}/(2\pi)^2=0.06 \, {\rm ps}^{2}$ as typical numbers for InAs/GaAs
 QDs~\cite{JiaoJPC2008,InAsParamRef}.
With these parameters, then the phonon-induced renormalization contribution,
$\langle  B\rangle=0.84$, at $T=10~$K.
With the polaron master equation, we can readily calculate the cavity-mode
{\em incoherent}
 emission, from:
$S_{cav}(\omega)  \propto
{\rm lim}_{t\rightarrow \infty}{\rm Re}\{ \int^{\infty}_{0}dt
[\braket{{a}(t+\tau){a}^{\dagger}(t)} -
\braket{a(t+\tau)}\braket{a^\dagger(t)}]  e^{i(\omega_{L}-\omega) t}\}$.
The master equation is
numerically solved in a basis that can be truncated
at any arbitrary photon and exciton state, which enables us to compute
the {\em weak
excitation approximation} results (one quanta limit) and the regime
of multi-quanta cavity-QED.
Since we consider cw excitation,
and the  phonon dephasing timescales are very fast,
we can take $t\rightarrow \infty$ in Eq.~(\ref{sec3eq3});
we have checked this approximation to be rigorously valid which confirms
that
that phonon processes occur at timescales substantially faster than the system dynamics.
The two-time correlation function
is obtained from the quantum regression formula~\cite{CarmichaelBook1}.

\begin{figure}[t!]
\centering
\includegraphics[trim = 10mm 16mm 20mm 12mm, clip, width=8.7cm ]{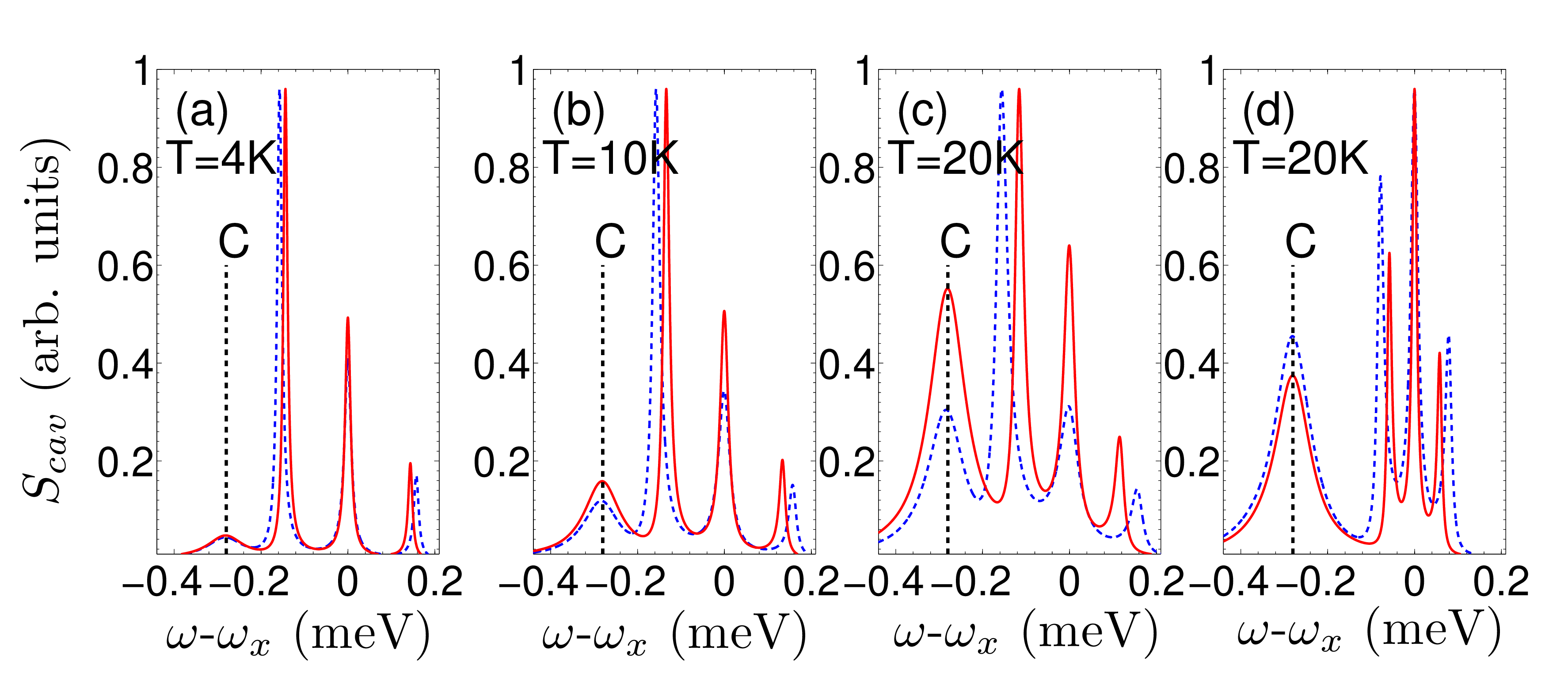}
\caption{(Color online)  Cavity-emitted fluorescence spectra for the one-phonon model (blue-dashed) and the full polaron theory (red-solid).
The cw Rabi drive strength $\Omega=0.05~$meV, and  is resonant with the
exciton ($\delta_{xL}=0$).
The parameters were as follows: $g=0.02$~meV, $\Gamma^{\prime}=2~\mu$eV, $\Gamma_{x}=1~\mu$eV, $2\kappa=0.1~$meV and $\omega_c-\omega_x=0.28\,$meV.
\label{fig2}}
\end{figure}

\begin{figure}
\centering
\includegraphics[trim = 10mm 0mm 0mm 12mm, clip, width=9cm ]{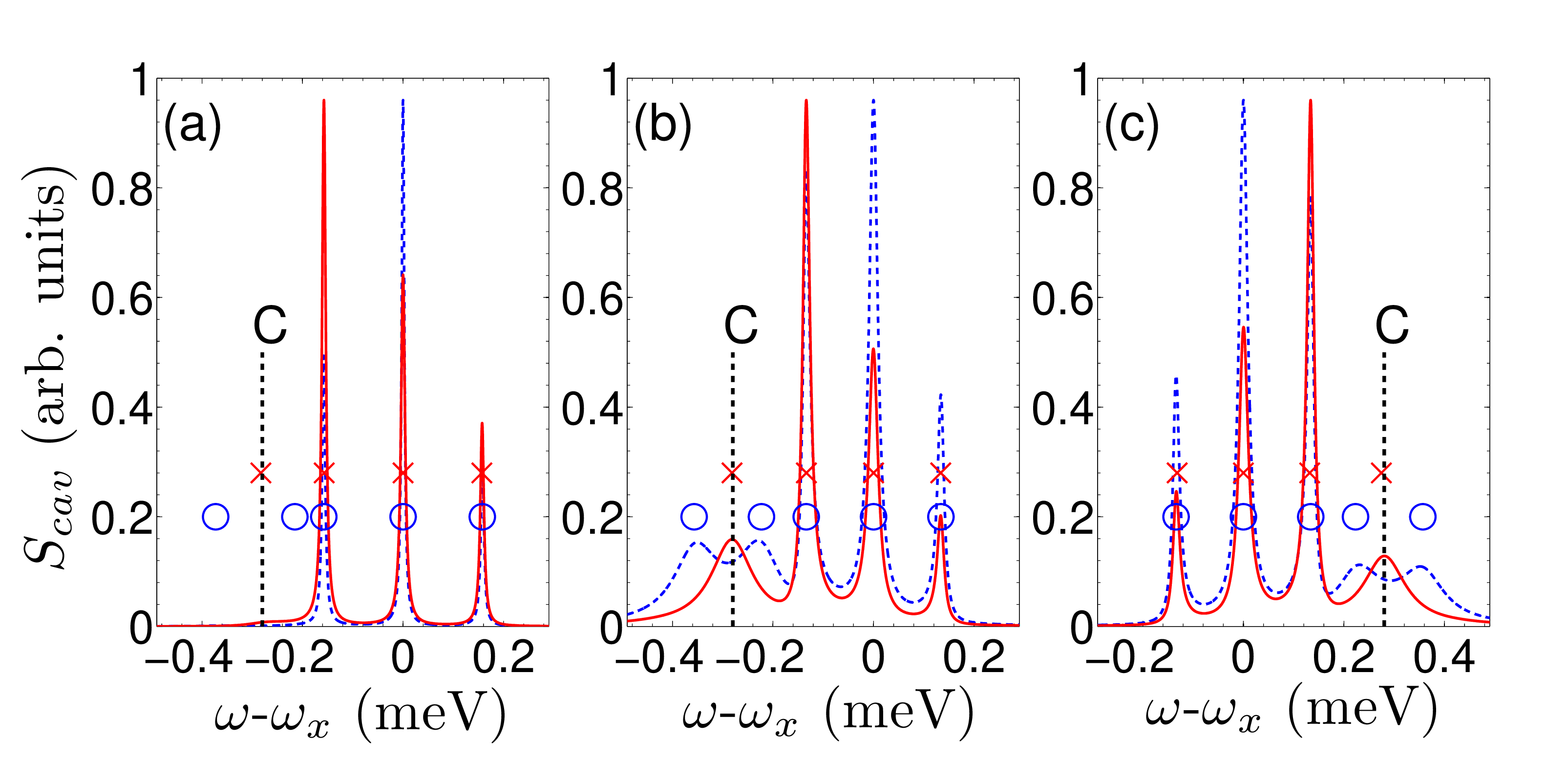}
\caption{(Color online) Cavity-emitted fluorescence  spectra with multi-photon processes (red-solid) compared to one-photon processes (blue-dashed) in the quantum dot cavity system.
We consider
phonons to all orders, and two different  dot-cavity coupling parameters (a) $g=0.02$~meV and (b) $g=0.05$~meV.
The cw Rabi drive and material parameters as the same as those used in Fig.~\ref{fig2}.
\label{fig3}}
\end{figure}
%

%
We
first study the need for both multi-phonons and multi-photons.
To be consistent with the recent Mollow triplet experiments on semiconductor micropillars~\cite{Ulrich:PLR2010},
we consider a detuned cavity mode with $\omega_x-\omega_c \approx 0.3\,$meV,
a cw resonantly-excited exciton, and
emitted spectra obtained from cavity-emission, $S_{cav}$.
 The radiative decay rate $\tilde{\Gamma}_{x}=1~\mu$~eV and the residual pure dephasing rate, $\Gamma^{\prime}=2~\mu$~eV;
 these are similar to those used by Ulrich {\em et al.}~\cite{Ulrich:PLR2010}, and between those
 measured in Ref.~\cite{ates1} and Ref.~\cite{flagg}.
  In Fig.~\ref{fig2}, we plot the cavity-emitted
  spectra for the one-phonon model and compare it with the full polaron theory (phonons to all orders). The one-phonon correlation function is obtained by expanding the phonon correlation functions $G_{g,u}(t)$,
   and $\langle  B\rangle$, to lowest order in the dot-phonon coupling.
   At low temperatures, the one-phonon expansion of the polaronic phonon correlation function is sufficient to describe the dynamics. However, with increasing temperatures, the one-phonon model starts to overestimate the effect of phonon coupling with respect to the polaronic description.
 This trend is in consistent with the pulse-excitation analysis of McCutcheon and Nazir~\cite{nazir2},
 for $T>30\,$K; however,
what is particularly surprising in our cw-cavity case
is that multi-phonon effects are important
at $T=10\,$K.

\begin{figure}[b!]
\centering
\includegraphics[trim = 12mm 0mm 0mm 10mm, clip, width=8.9cm]{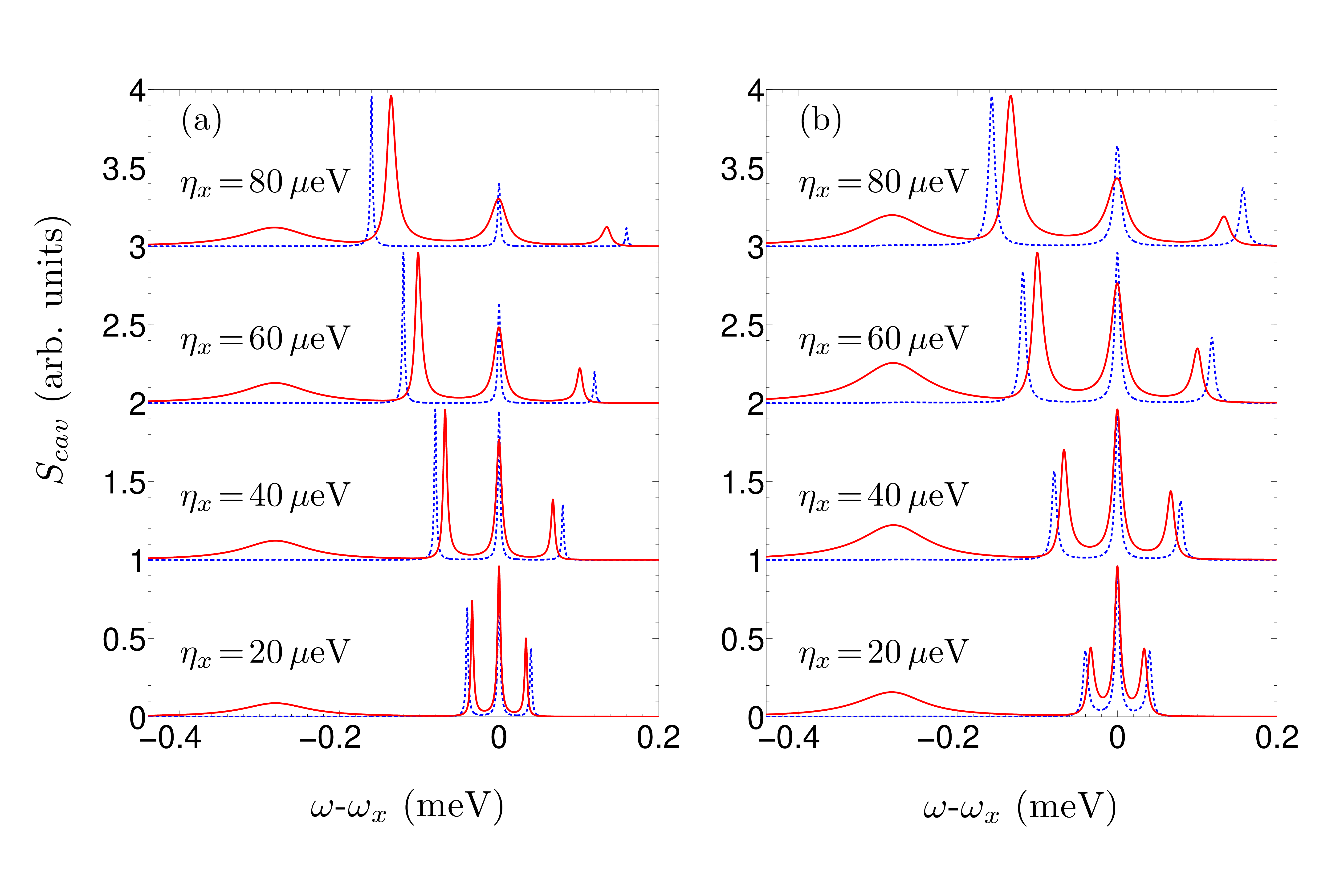}
\caption{(Color online) Cavity-emitted  fluorescence spectra of a quantum dot placed under cw
 excitation of the exciton (with $\delta_{x}=0$).
 We vary the Rabi frequency $\Omega$ from $0.01-0.05~$meV, and we
  consider two different quantum dot-cavity coupling parameters (a) $g=15$~$\mu$eV and (b) $g=50$~$\mu$eV. The blue-dashed plots correspond to the cases with no phonon coupling, while the red-solid line corresponds to the case where the quantum dot is coupled to a phonon reservoir at a temperature $T=10\,$K.
The other parameters are as in Fig.~\ref{fig2}.
\label{fig4}}
\end{figure}

  In Fig.~\ref{fig3} we  study the role of multi-photon processes in the QD cavity dynamics in the presence of phonons, for different values of dot-cavity coupling parameters. We find that a one-photon-correlation approximation fails in the present
 excitation regime, though a two-photon-correlation basis was found to be sufficient for all the
 calculations that follow.  This confirms that the theory
 must include multi-phonon and multi-photon effects, {\em even for a detuned cavity system at $T=10\,$K}.
 With photons and phonons included, we
 recognize
  two remarkable features in  the cavity-emitted spectra.
First, there is a strong asymmetry for the on-resonance Mollow triplet,
which is primarily due to the off-resonant cavity.
Second, there is a pronounced cavity feeding effect, where the cavity mode
is excited via phonon interactions, which is
a similar effect to incoherently excited QD-cavity systems \cite{HennessyNature:2007,KaniberPRB:2008,SufczynskiPRL:2009,JiaoJPC2008,ota,HohenesterPRB:2009,InAsParamRef},
but much stronger in this coherent excitation regime.
The
asymmetries of the Mollow triplet are attributed to the subtle interplay of the cavity-photon bath, phonon bath and pure dephasing.

%

\begin{figure}[t!]
\includegraphics[trim = 6mm 0mm 0mm 22mm, clip, width=8.6cm]{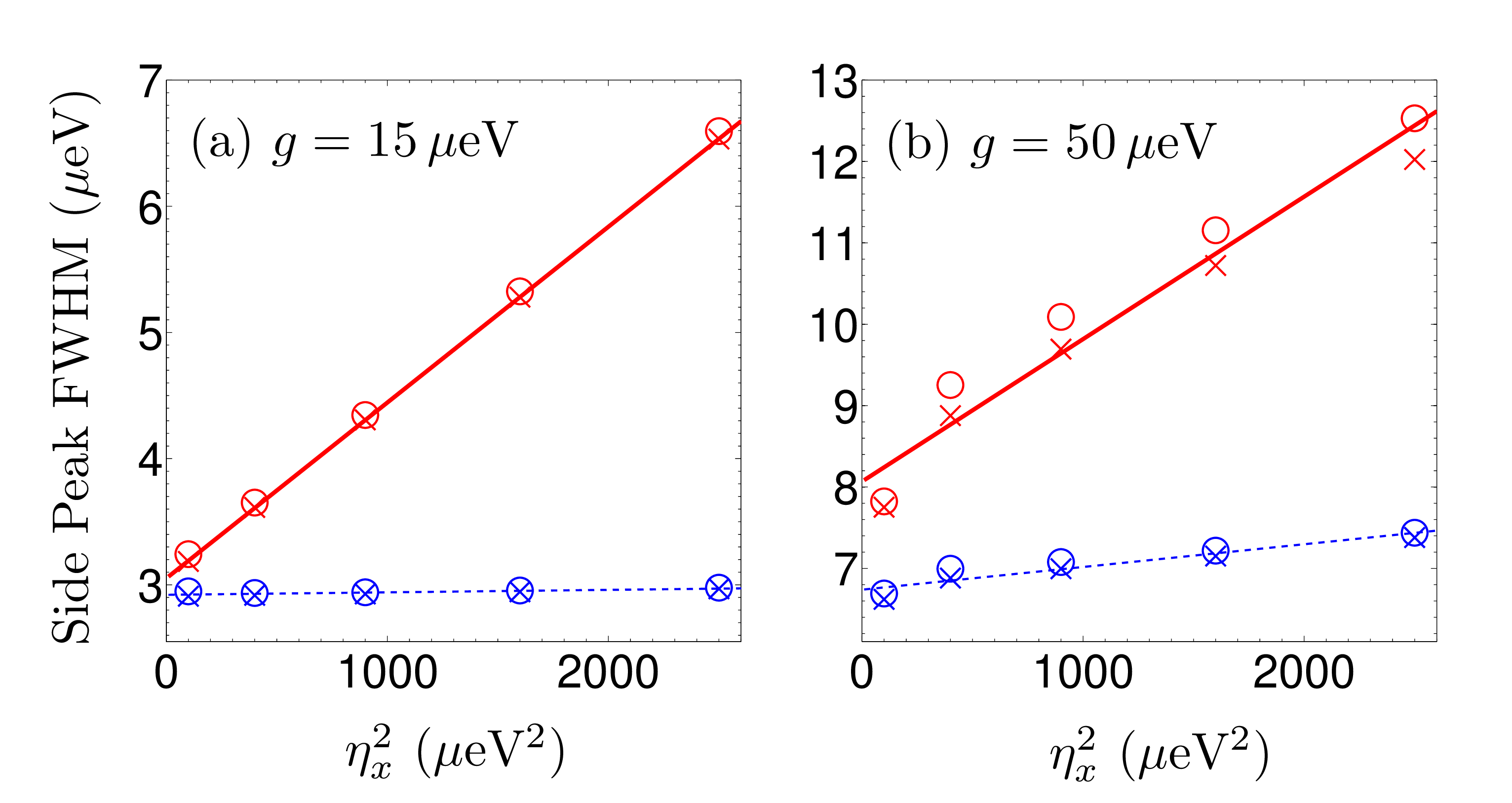}
\vspace{-0.7cm}
\caption{(Color online)  FWHM of the Mollow sidebands as a function of square of the Rabi frequency ($\Omega^{2}$) extracted by a numerical fit from Fig.~\ref{fig4} in the presence of phonon coupling (red-solid) and with no phonon coupling (blue-dashed). The linear fit is a signature of excitation-induced dephasing due to the underlying phonon coupling to the quantum dot.
\label{fig5}}
\end{figure}

In Fig.~\ref{fig4}, we show the cavity-emitted spectrum of the QD exciton under resonant
 excitation conditions ($\delta_{xL}=0$) as a function of cw excitation Rabi frequency, for  $g=15$~ $\mu$eV and $g=50$~$\mu$eV. The spectra are normalized to unity, and, for clarity, the spectra have been shifted vertically as a function
of Rabi field strength. We show
spectra with and without phonon coupling, where the former
assumes a
phonon reservoir
in thermal equilibrium at $T=10\,$K.
As expected, we clearly find increasing splitting of the Mollow triplet with increasing Rabi frequency $\Omega$. We also note that the location of the sidebands changes due to the renormalization of the Rabi frequency in the presence of phonon coupling which reduces the effective Rabi frequency; the Mollow splitting is slightly reduced as a result and the triplet resonances
are significantly broader. In Fig.~\ref{fig5}, we plot the corresponding
numerically-extracted FWHM (full-width half-maximum) values as a function of the square of the Rabi frequency. We find a very good linear fit to the data which suggests that Mollow triplets are broadened with increasing $\Omega$, in agreement with recent experiments \cite{Ulrich:PLR2010}.
However, for the case of $g=50$~ $\mu$eV, the dependence is likely nonlinear for low values of $\Omega$, and in general dephasing results from the interplay
of both phonon bath coupling and cavity photon bath coupling. In addition, we obtain
rich asymmetries of the Mollow triplet and substantial cavity feeding -- two pronounced
effects that are {\em unique} to the QD-cavity system.

In summary, we have presented a
theoretical and numerical study of the cavity-emitted
resonance fluorescence of a QD placed inside a high $Q$ cavity. We developed
 a polaron master equation formalism based on the time-convolutionless approach which enabled us to study the dynamical properties of the reduced density matrix of the QD-cavity system  in response to both an external cw laser field and a bath of thermalized LA phonons.
We applied this theory to study the
 influence of the Rabi frequency on
 cavity-emitted spectra and find strong numerical evidence of excitation-induced dephasing.
 Both multi-phonon and multi-photon effects are shown to be important and our
 general findings are in excellent agreement with the very recent experiments of Ulrich {\em et al.}~\cite{Ulrich:PLR2010}.


This work was supported by the
National Sciences and Engineering Research Council of
Canada and the Canadian Foundation for Innovation.
We gratefully acknowledge S. Reitzenstein and H. Carmichael for
useful discussions, and  S. Ulrich, S. Reitzenstein, and P. Michler
for sharing their experimental results~\cite{Ulrich:PLR2010}
prior to publication.


\end{document}